\documentclass[aps,twocolumn,prx]{revtex4-1}
\usepackage{graphicx,amsmath, amsfonts,bm, color}

\usepackage{xcolor,hyperref}
\hypersetup{
   colorlinks,
   linkcolor={blue!50!black},
   citecolor={blue!50!black},
   urlcolor={blue!80!black}
}

\usepackage{enumitem}
\usepackage{epsfig, amsmath}
\usepackage{bm}
\usepackage{soul}
\DeclareMathAlphabet{\mathitbf}{OML}{cmm}{b}{it}
\newcommand{\dv}{\mathitbf d}
\newcommand{\xv}{\mathitbf x}
\newcommand{\uv}{\mathitbf u}
\newcommand{\zv}{\mathitbf z}
\newcommand{\dbar}{{\,\mathchar'26\mkern-12mu d}}
\newcommand{\calBold}[1]{\mbox{\boldmath${\cal #1}$}}
\newcommand{\mathBold}[1]{\mbox{\boldmath$#1$}}

\begin{document}

\title{Pinching a glass reveals key properties of its soft spots}
\author{Corrado Rainone$^1$}
\author{Eran Bouchbinder$^2$}
\author{Edan Lerner$^1$}
\thanks{e.lerner@uva.nl}
\affiliation{$^1$Institute for Theoretical Physics, University of Amsterdam, Science Park 904, 1098 XH Amsterdam, The Netherlands \\ $^2$Chemical and Biological Physics Department, Weizmann Institute of Science, Rehovot 7610001, Israel }

\begin{abstract}
It is now well established that glasses feature quasilocalized nonphononic excitations --- coined ``soft spots''---, which follow a universal $\omega^4$ density of states in the limit of low frequencies $\omega$. All glass-specific properties, such as the dependence on the preparation protocol or composition, are encapsulated in the non-universal prefactor of the universal $\omega^4$ law. The prefactor, however, is a composite quantity that incorporates information both about the number of quasilocalized nonphononic excitations and their characteristic stiffness, in an apparently inseparable manner. We show that by pinching a glass, i.e.~by probing its response to force dipoles, one can disentangle and independently extract these two fundamental pieces of physical information. This analysis reveals that the number of quasilocalized nonphononic excitations follows a Boltzmann-like law in terms of the parent temperature from which the glass is quenched. The latter, sometimes termed the fictive (or effective) temperature, plays important roles in non-equilibrium thermodynamic approaches to the relaxation, flow and deformation of glasses. The analysis also shows that the characteristic stiffness of quasilocalized nonphononic excitations can be related to their characteristic size, a long sought-for length scale. These results show that important physical information, which is relevant for various key questions in glass physics, can be obtained through pinching a glass.
\end{abstract}

\maketitle

\section{Introduction}

Understanding the micromechanical, statistical and thermodynamic properties of soft nonphononic excitations in structural glasses remains one of the outstanding challenges in glass physics, despite decades of intensive research~\cite{soft_potential_model_1987,soft_potential_model_1991,Gurevich2003, Gurevich2007,matthieu_thesis,mw_EM_epl,Schirmacher_2006,Schirmacher_prl_2007,Marruzzo2013, eric_boson_peak_emt, silvio,Beltukov2015,modes_prl,modes_prl_2018,cge_paper,lisa_random_matrix_2019,parisi_mean_field_w4,h_ikeda_pre_2019, ikeda_2019_dipole}. Soft nonphononic excitations are believed to give rise to a broad range of glassy phenomena, many of which are still poorly understood; some noteworthy examples include the universal thermodynamic and transport properties of glasses at temperatures of 10K and lower \cite{Zeller_and_Pohl_prb_1971,Anderson, Phillips, soft_potential_model_1991,Gurevich2007}, the low-temperature yielding transition in which a mechanically-loaded brittle glass fails via the formation of highly localized bands of plastic strain \cite{Schuh_review_2007,falk_review_2016}, and anomalous, non-Rayleigh wave attenuation rates \cite{lemaitre_tanaka_2016, Ikeda_scattering_2018, scattering_jcp}.

Computational studies have been invaluable in advancing our knowledge about the statistical and mechanical properties of soft glassy excitations, and in revealing the essential roles that these excitations play in various glassy phenomena. Dating back to the early 1990's, Schober and Laird were the first to reveal the existence of soft spots in the form of low-frequency, quasilocalized vibrational modes in a model computer glass~\cite{Schober_Laird_numerics_PRL,Schober_Laird_numerics_PRB}. Soon later, Schober and coworkers showed that relaxation events deep in the glassy state exhibit patterns that resemble quasilocalized modes~\cite{schober1993_numerics}, suggesting a link between soft glassy structures and dynamics. In an important subsequent work~\cite{widmer2008irreversible}, this link was further strengthened by showing that relaxational dynamics in supercooled liquids strongly correlates with quasilocalized low-frequency vibrational modes measured in underlying inherent states. Some years later, it was shown that plastic activity in model structural glasses and soft-sphere packings is intimately linked to nonphononic low-frequency modes~\cite{tanguy2010,manning2011,rottler_normal_modes}.

It was, however, only recently that the universal statistical and structural properties of soft quasilocalized modes in glasses were revealed, first in a Heisenberg spin glass in a random field \cite{parisi_spin_glass}, and later in model structural glasses \cite{modes_prl, SciPost2016,ikeda_pnas, modes_prl_2018,LB_modes_2019}. It is now well accepted that the density of nonphononic quasilocalized modes of frequency $\omega$ grows from zero (i.e.~without a gap) as $\omega^4$, independently of microscopic details~\cite{modes_prl}, preparation protocol \cite{cge_paper}, or spatial dimension \cite{modes_prl_2018}. Importantly, as shown in~\cite{LB_modes_2019} and demonstrated again in this work, the $\omega^4$ distribution of quasilocalized modes persists even in inherent states that underlie very deeply supercooled states, i.e.~in stable computer glasses whose stability is comparable to conventional laboratory glasses. Furthermore, soft quasilocalized modes have been shown to generically feature a disordered core of linear size of a few particle spacings, decorated by long-range Eshelby-like displacement fields whose amplitude decays as $r^{1-\dbar}$ at a distance $r$ from the core, in $\dbar$ spatial dimensions \cite{modes_prl, modes_prl_2018}.
\begin{figure}[ht!]
\centering
\begin{tabular}{ccc}
\includegraphics[width=0.90\linewidth]{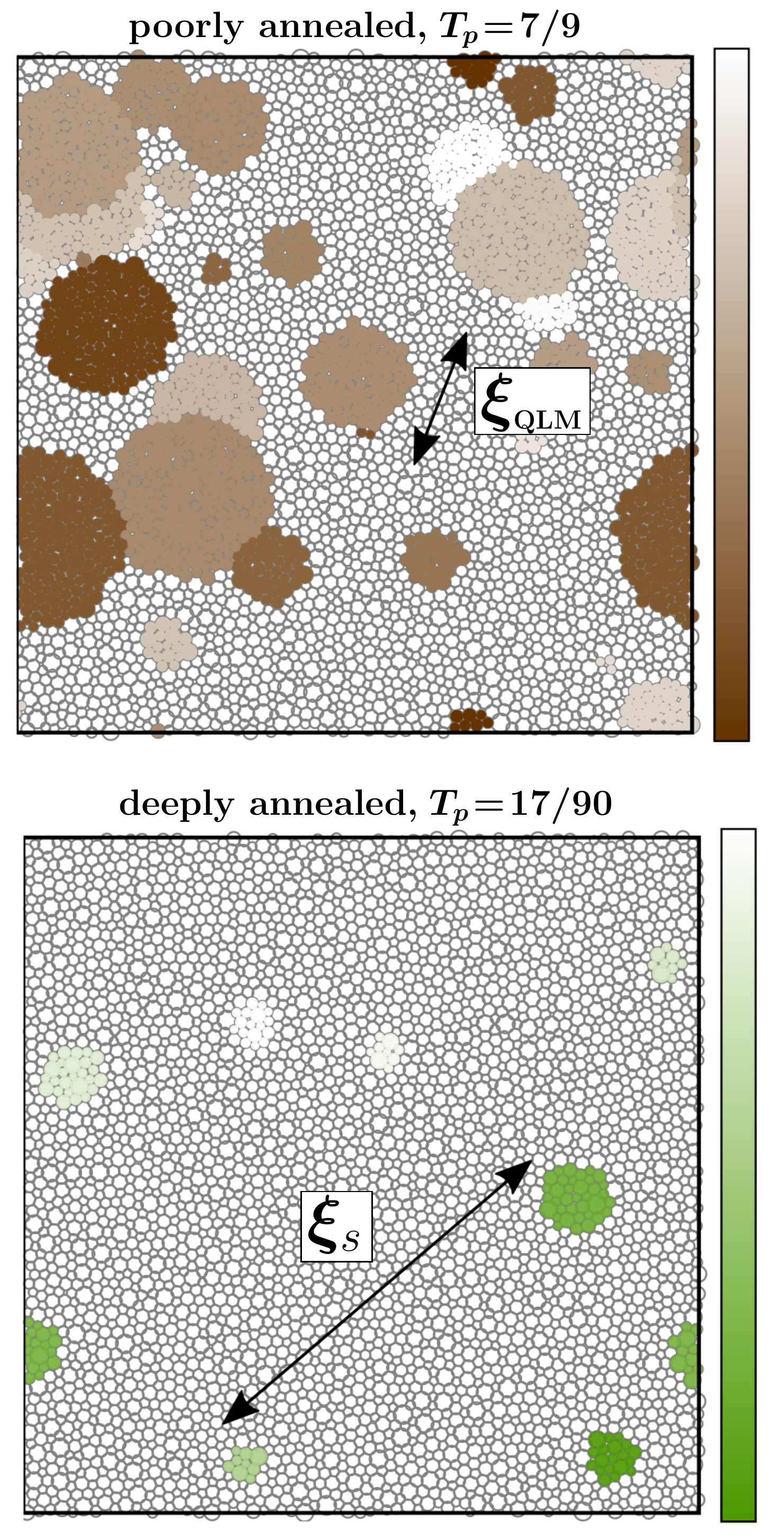}
\end{tabular}
\caption{A graphical representation of the population of QLMs in poorly annealed (upper panel) and deeply annealed (bottom panel) 2D computer glasses. Each blob represents a QLM, its size is proportional to our estimation of the mode's core size $\xi_{\mbox{\tiny QLM}}$, and the color code represents the mode's frequency, decreasing from bright to dark; upper (bottom) panel color code range is [0.18,0.42] $\big($[0.54,0.74]$\big)$, expressed in terms of $c_\infty/a_0$, with $c_\infty$ being the high-$T_p$ shear wave speed, and $a_0$ is the interparticle distance. The typical distance between QLMs, $\xi_s$, is also marked. Details of the calculation can be found in \cite{in_prep} and in the SI. Note that the deeply annealed case shown in the bottom panel might be representative of laboratory molecular or metallic glasses.}
\label{fig:Fig1}
\end{figure}

The key challenge in revealing the statistical, structural and energetic properties of soft quasilocalized modes --- to be termed hereafter QLMs --- in computer investigations lies in the abundance of spatially-extended low-frequency phonons in structural glasses~\cite{SciPost2016, phonon_widths}. These phononic excitations hybridize with quasilocalized excitations, as pointed out decades ago by Schober and Oligschleger~\cite{SchoberOligschleger1996}. These hybridization processes hinder the accessibility of crucial information regarding characteristic length and frequency scales of QLMs, and regarding their prevalence.

While promising attempts to overcome the aforementioned hybridization issues have been put forward~\cite{SchoberOligschleger1996,SciPost2016,lte_pnas,manning_defects}, a complete statistical-mechanical picture of QLMs is still lacking. In particular, recent work has revealed that annealing processes affect QLMs in three ways: firstly, the number of QLMs appears to decrease upon deeper annealing, i.e.~they are depleted, as first pointed out in~\cite{protocol_prerc, cge_paper}. Secondly, the core size of QLMs, $\xi_{\mbox{\tiny QLM}}$, was shown to decrease with deeper annealing~\cite{modes_prl, inst_note}. Lastly, in~\cite{modes_prl,cge_paper} it was shown that the characteristic frequencies of QLMs also increase upon deeper annealing, i.e.~they stiffen, in addition to their depletion. These three effects, and other concepts discussed below, are graphically illustrated in Fig.~\ref{fig:Fig1}.

In this work, we investigate the effect of very deep supercooling/annealing on the statistical, structural and energetic properties of QLMs in a model computer glass (see Methods section for details). First, we explain why information regarding the number of QLMs \emph{cannot} typically be obtained from the universal vibrational density of states (vDOS) of QLMs alone. Instead, we show that the vDOS grants access to a composite physical observable, which encodes information regarding both the characteristic frequency scale of QLMs, $\omega_g$, and their number, ${\cal N}$. Then, following recent suggestions~\cite{new_variational_argument_epl_2016,cge_paper}, we use the average response of the glass to a local pinch --- more formally, we use the bulk-average response of a glass to force dipoles --- as a measure of $\omega_g$. This assumption, in turn, allows us to \emph{quantitatively} disentangle the processes of annealing-induced stiffening of QLMs from their annealing-induced depletion.

Interestingly, this analysis reveals that ${\cal N}$ follows an equilibrium-like Boltzmann relation ${\cal N}\!\propto\!\exp\!\Big(\!\!\!-\hspace{-0.07cm}\frac{E_{_{\mbox{\tiny QLM}}}}{k_{\!_B} T_p}\!\Big)$, with $T_p$ denoting the parent temperature from which glassy states are quenched, $k_{\!_B}$ is Boltzmann's constant, and $E_{_{\mbox{\tiny QLM}}}$ is the energetic cost of creating a QLM. That is, our results indicate that QLMs behave as ``quasiparticles'' whose number is determined by equilibrium statistical thermodynamics at the parent equilibrium temperature $T_p$, and that this number is preserved when the glass goes out-of-equilibrium during a quick quench to a temperature much smaller than $T_p$. The QLMs thus appear to correspond to configurational degrees of freedom that carry memory of the equilibrium state at $T_p$, deep into the non-equilibrium glassy state, and in this sense $T_p$ has a clear thermodynamic interpretation as a non-equilibrium temperature. This physical picture has been for quite some time the cornerstone of the non-equilibrium thermodynamic Shear-Transformation-Zones (STZs) theory of glass deformation~\cite{Bouchbinder2009b, Bouchbinder2009c, Falk2011}, where $T_p$ is termed a fictive/effective/configurational temperature, once QLMs are identified with STZs, i.e.~with glassy ``flow defects''~\cite{falk_langer_stz}.

Furthermore, we show that $\omega_g$ can be used to define a length that appears to match the independently determined core size of QLMs, argued to mark the crossover between the disorder-dominated elastic response of glasses at the mesoscale, and the continuum-like elastic response at the macroscale~\cite{breakdown}. Taken together, these results show that important physical information, which is relevant for various key questions concerning the formation, relaxation and flow of glasses, can be obtained through pinching a glass.

\section{The QLMs depletion versus stiffening conundrum}

It is now established that the vDOS of QLMs, ${\cal D}(\omega)$, follows a universal gapless law~\cite{modes_prl, SciPost2016,ikeda_pnas, modes_prl_2018,LB_modes_2019}
\begin{equation}
\label{foo00}
{\cal D}(\omega) = A_g\,\omega^4 \quad\quad\hbox{for}\quad\quad 0 \le \omega \le \omega_g \ ,
\end{equation}
where $\omega_g$ is the upper cutoff of this scaling regime and the prefactor $A_g$ is extensively discussed below. The $\omega^4$ law has been rationalized by various models~\cite{soft_potential_model_1987,soft_potential_model_1991, Gurevich2003, Gurevich2007, parisi_mean_field_w4, lisa_random_matrix_2019, h_ikeda_pre_2019, ikeda_2019_dipole}, and is known to be intimately related to the existence of frustration-induced internal stresses in glasses~\cite{inst_note}, but its theoretical foundations are not yet fully developed. The prefactor $A_g$ in~\eqref{foo00} (denoted by $A_4$ in~\cite{LB_modes_2019, h_ikeda_pre_2019, mw_thermal_origin}) is a non-universal quantity that encodes information about a particular glassy state, most notably its composition (constituent elements, interaction potential etc.) and its preparation protocol~\cite{protocol_prerc,cge_paper,LB_modes_2019}. The ultimate goal of this work is to explore the physical information encapsulated in $A_g$ and its dependence on the glass preparation protocol.
\begin{figure}[ht]
\centering
\begin{tabular}{ccc}
\includegraphics[width=1.00\linewidth]{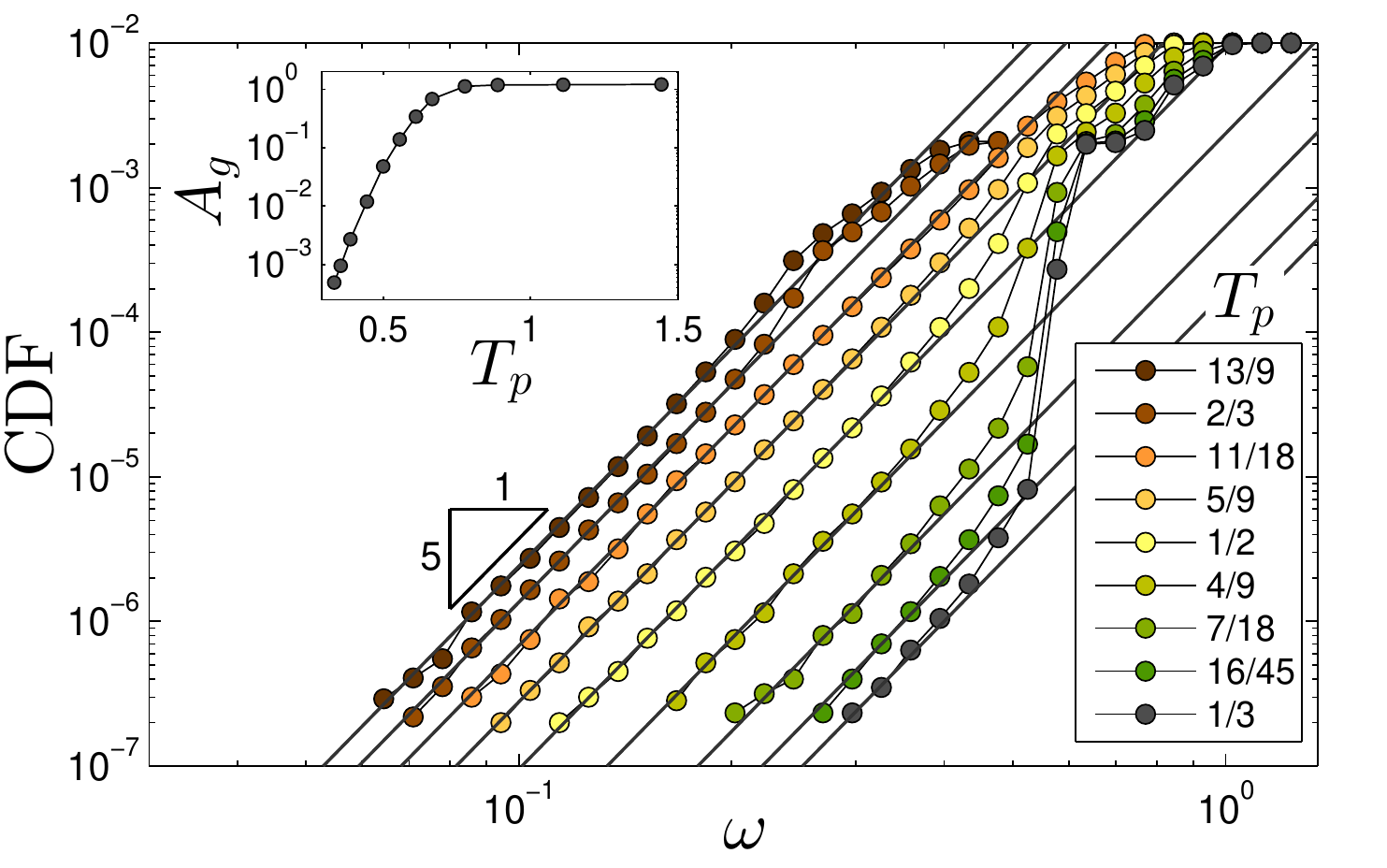}
\end{tabular}
\caption{Cumulative density of states $\mbox{CDF}\!\equiv\!\int_0^\omega {\cal D}(\omega')d\omega'$ for various parent temperatures $T_p$ (see values in the legend). Here we present data measured in $10000$ glassy samples of $N\!=\!2000$ particles for $T_p\!\le\!11/18$ and $2000$ samples of $N\!=\!16000$ particles for $T_p\!>\!11/18$. Inset: the prefactors $A_g$ vs.~$T_p$, see text for discussion.}
\label{fig:Fig2}
\end{figure}

In Fig.~\ref{fig:Fig2}, we plot the cumulative vDOS calculated for glassy samples rapidly quenched from parent equilibrium temperatures $T_p$ (as appears in the figure legend) to zero temperature. The system size is chosen so as to avoid hybridization with phonons at the lowest frequencies, as explained in~\cite{modes_prl}. The figure shows, in agreement with~\cite{LB_modes_2019}, that the $\omega^4$ scaling persists all the way down to the deepest supercooled states accessible to us, $T_p\!=\!1/3$ (the units used to report $T_p$ are defined below). The inset shows that the prefactor $A_g$ varies by nearly $3$ orders of magnitude in the simulated $T_p$ range. The huge variability of $A_g$ with the preparation protocol, here quantified by the parent equilibrium temperature $T_p$, indicates dramatic changes in the resulting glassy states, despite the fact that all of them follow the universal $\omega^4$ law.

What physics is encapsulated in $A_g$? To start addressing this question, let us first consider the dimensions of $A_g$. When ${\cal D}(\omega)$ is integrated over the frequency range in which~\eqref{foo00} is valid, i.e.~in the range $0\!\le\!\omega\!\le\!\omega_g$, one obtains an estimate for the total number of QLMs, ${\cal N}$. Consequently, $A_g$ has the dimensions of an inverse frequency to the fifth power, where the dimensionless prefactor is proportional to ${\cal N}$. Since ${\cal D}(\omega)$ follows a power-law, i.e.~it is scale-free in the range $0\!\le\!\omega\!\le\!\omega_g$, the only possible frequency scale that can appear in it is the upper cutoff $\omega_g$. Hence, we expect to have $A_g\!\sim\!{\cal N}\omega_g^{-5}$, which implies that~\eqref{foo00} should be rewritten as
\begin{equation}
\label{foo01}
{\cal D}(\omega) \sim {\cal N}\, \omega_g^{-5}\omega^4 \quad\quad\hbox{for}\quad\quad 0 \le \omega \le \omega_g \ .
\end{equation}

We would like to note the analogy, and the fundamental difference, between~\eqref{foo01} and Debye's vDOS of (acoustic) phononic excitations in crystalline solids~\cite{kittel2005introduction}. The latter takes the form $D(\omega)\! =\! A_D\,\omega^2$ (in three dimensions), with $A_D\! = \! 9N/\omega_D^3$, where $\omega_D$ is Debye's frequency and $N$ is the number of particles. The integral over $D(\omega)$ in the range $0\!\le\!\omega\!\le\!\omega_D$ equals the number of degrees of freedom in the system, $3N$. The analogy between Debye's vDOS and the glassy vDOS in~\eqref{foo01}, and between $\omega_g$ and Debye's frequency, is evident. Yet, there is a crucial difference between the two cases; in Debye's theory the number of phononic excitations is a priori known to equal the number of degrees of freedom $3N$ (in fact, $\omega_D$ is precisely defined so as to ensure the latter). In the glassy case, however, there is neither an a priori constraint on the number of QLMs ${\cal N}$, nor on the upper frequency cutoff $\omega_g$ (the total number of vibrational modes, both glassy and phononic, is of course still determined by the total number of degrees of freedom, but there is no a priori constraint on the fraction of QLMs out of the total number of vibrational modes). Hence, ${\cal N}$ and $\omega_g$ should be treated as independent quantities that can feature different dependencies on the glass history (preparation protocol).
\begin{figure}[ht]
\centering
\begin{tabular}{ccc}
\includegraphics[width=1.00\linewidth]{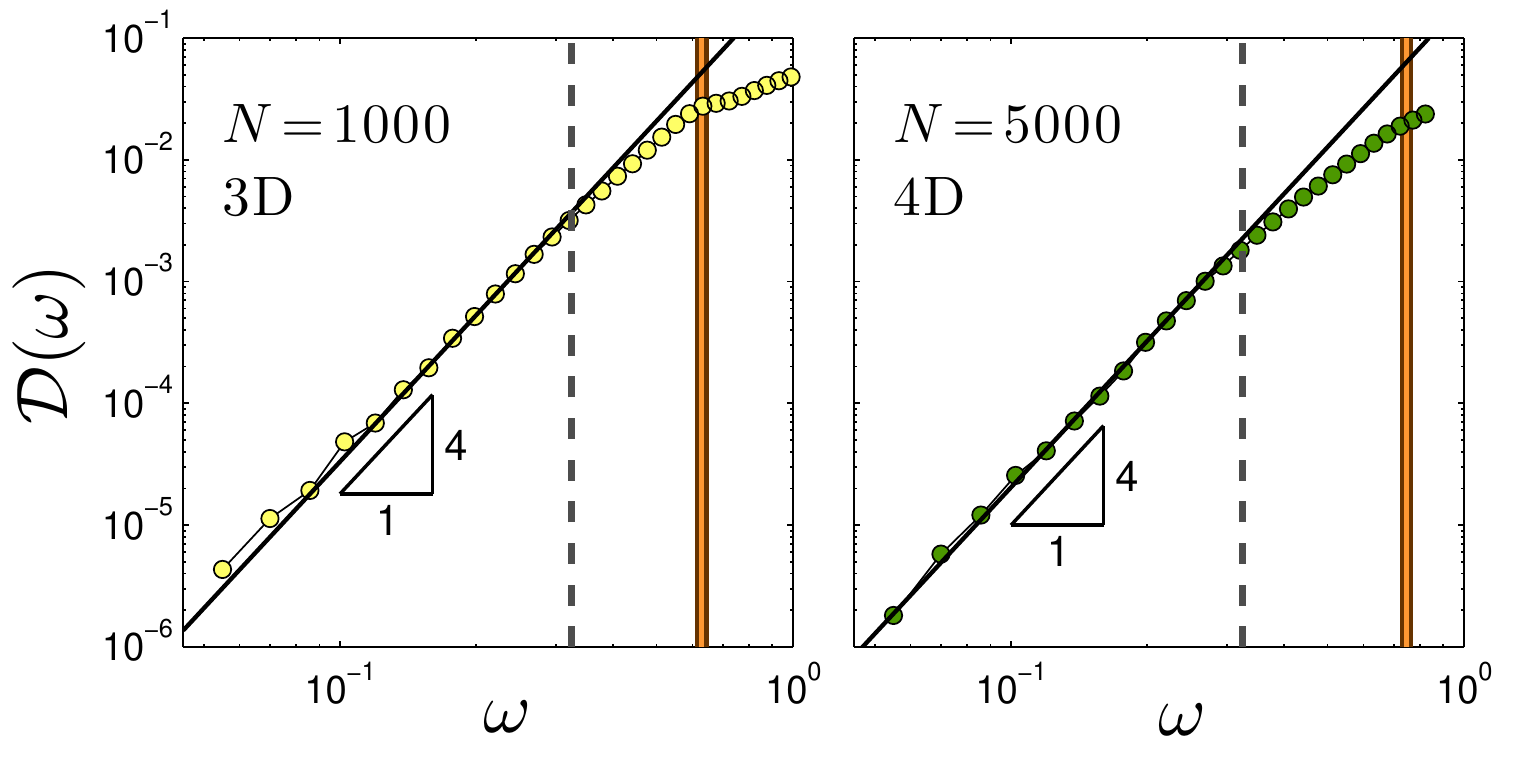}
\end{tabular}
\caption{The vDOS ${\cal D}(\omega)$ of small glassy samples (the number of particles $N$ is specified in each panel) in three dimensions (3D, left) and four dimensions (4D, right), obtained by a rapid quench in~\cite{modes_prl_2018}. The data are adapted from Fig.~2b-c of~\cite{modes_prl_2018}, where frequencies are normalized as detailed in the SI. The vertical continuous lines indicate the position of the first phonon band, whereas the dashed lines mark the breakdown of the $\omega^4$ scaling regime.}
\label{fig:Fig3}
\end{figure}

In order to disentangle the number of QLMs (${\cal N}$) and their characteristic frequency ($\omega_g$) contributions to $A_g\!\sim\!{\cal N}\omega_g^{-5}$, one needs to estimate one of them, i.e.~either ${\cal N}$ or $\omega_g$, independently of $A_g$. In principle, as the characteristic frequency $\omega_g$ represents the upper cutoff on the $\omega^4$ scaling regime (as explained above), one can try to estimate it through the deviation from the universal $\omega^4$ law. This has been, in fact, demonstrated in~\cite{modes_prl_2018} for rapidly quenched glassy samples in a narrow range of system sizes, in three and four dimensions. Some of the data appearing in Fig.~2b-c of~\cite{modes_prl_2018} are reproduced here in Fig.~\ref{fig:Fig3}, where the lowest phononic band is shown in orange in each panel. It is observed that in these examples, the vDOS deviates from the $\omega^4$ scaling at a frequency smaller than the lowest phonon frequency, which can be identified with $\omega_g$ (marked by the vertical dashed lines).

In general, though, the lowest phonon frequency is in fact smaller than $\omega_g$, which obscures the identification of the latter due to hybridizations~\cite{phonon_widths}. Indeed, in Fig.~\ref{fig:Fig2} it is observed that as $T_p$ decreases, the lowest phonon band pushes the vDOS upwards in the middle of the scaling regime, disallowing to extract $\omega_g$. Hence, we conclude that the vDOS alone does not allow one to distinguish between changes in the number of QLMs (e.g.~a decrease, i.e.~depletion) and in their characteristic frequency (e.g.~an increase, i.e.~stiffening). How to disentangle the ${\cal N}$ and $\omega_g$ dependence of $A_g$, and the possible depletion and stiffening of QLMs associated with them, is the question we address next.

\section{Estimating QLMs frequency scale by pinching a glass}

The previous discussion showed that the $T_p$ dependence of $A_g\!\sim\!{\cal N}\omega_g^{-5}$ cannot be readily used to extract the $T_p$ dependence of ${\cal N}$ and $\omega_g$ separately. Consequently, one needs additional physical input in order to disentangle the two quantities. Here we follow the suggestion put forward in~\cite{cge_paper} that the characteristic frequency $\omega_g$ of QLMs can be probed through pinching a glass. Formally, by pinching we mean applying a force dipole $\dv^{(ij)}$ to a pair of interacting particles $i,j$ in a glassy sample. The displacement response to $\dv^{(ij)}$, which was shown to closely resemble the spatial pattern of QLMs~\cite{cge_paper}, can be associated with a frequency $\omega_g^{(ij)}$ (see additional details in the SI). By averaging $\omega_g^{(ij)}$ over many interacting pairs $i,j$ in a glassy sample, one obtains a characteristic frequency scale, which was proposed to represent $\omega_g$. This suggestion was discussed at length and tested, under various circumstances, in~\cite{cge_paper}; here we follow it, i.e.~assume that the $T_p$ dependence of the dipole response is proportional to $\omega_g(T_p)$. The remainder of the paper is devoted to exploring the implications of this assumption.
\begin{figure}[ht]
\centering
\begin{tabular}{ccc}
\includegraphics[width=1.00\linewidth]{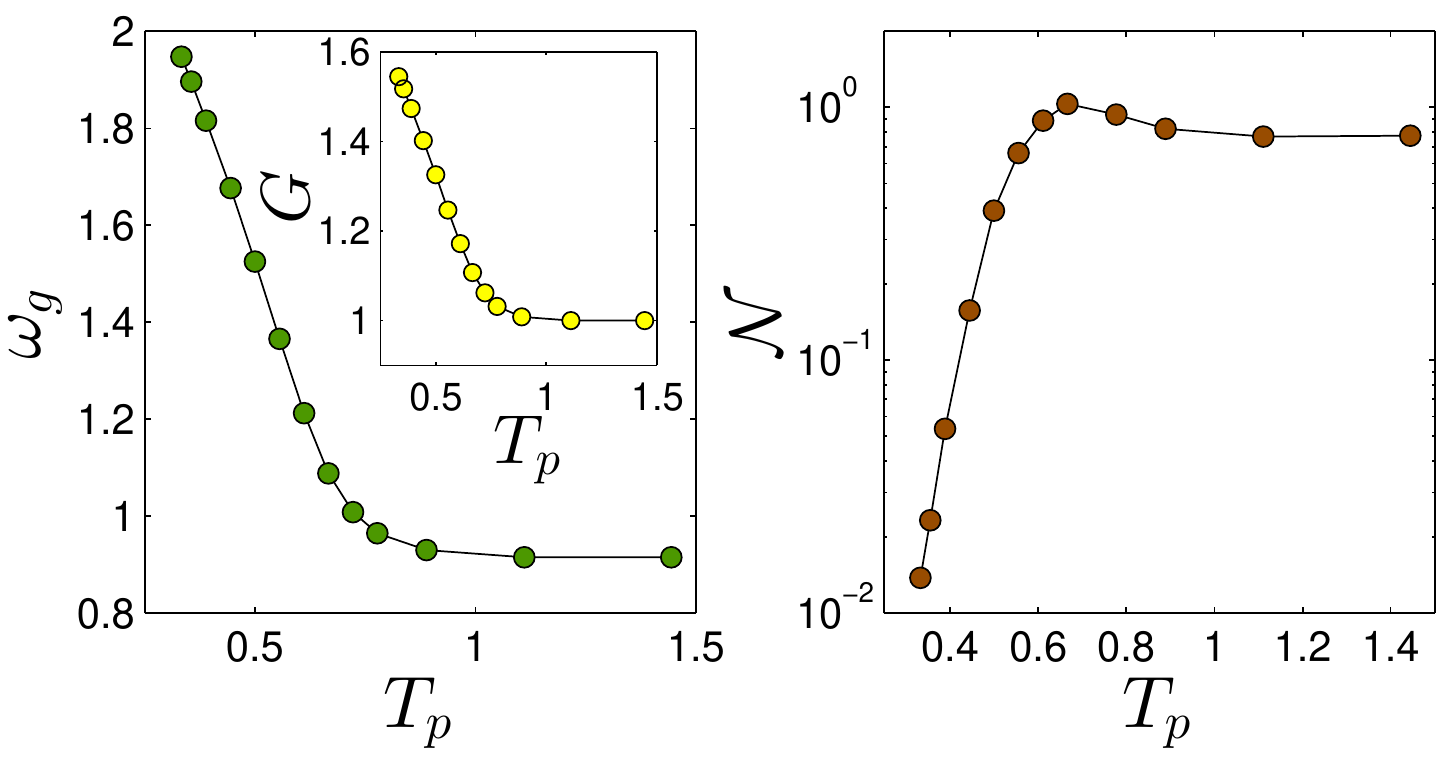}
\end{tabular}
\caption{(left) The characteristic frequency $\omega_g$ of quasilocalized modes, estimated by the pinching procedure discussed in the text, plotted vs.~the parent temperature $T_p$. Inset: the sample-to-sample mean athermal shear modulus, $G$, plotted against $T_p$. (right) ${\cal N}$ is proportional to the number of quasilocalized modes, and is plotted here against $T_p$.}
\label{fig:Fig4}
\end{figure}

In Fig.~\ref{fig:Fig4}a we plot the characteristic frequency $\omega_g$ vs.~the parent temperature $T_p$, where $\omega_g(T_p)$ is estimated by the pinching procedure just described. It is observed that $\omega_g$ varies by nearly a factor of $2$ at low parent temperatures $T_p$ and reaches a plateau at higher $T_p$. We further find that the sample-to-sample mean athermal shear modulus, $G$, shown in the inset, also plateaus at the same $T_p$ as $\omega_g$ does. Consequently, in what follows we conveniently express temperatures in terms of the onset temperature $T_{\mbox{\tiny onset}}$ of the high-$T_p$ plateaus of $G$ and $\omega_g$.

We conclude that, in the $T_p$ range considered here, QLMs appear to stiffen by a factor of approximately 2 with decreasing $T_p$. Interestingly, in~\cite{LB_modes_2019} it was reported that the boson peak frequency $\omega_{\mbox{\tiny BP}}$ varies by approximately a factor of 2 over a similar range of $T_p$, suggesting that $\omega_{\mbox{\tiny BP}}$ and $\omega_g$ might be related. In~\cite{new_variational_argument_epl_2016}, a similar proposition was put forward in the context of the unjamming transition~\cite{ohern2003,liu_review,van_hecke_review}, where it was argued that the renowned `unjamming' frequency scale $\omega_*$~\cite{ohern2003, matthieu_thesis} can be extracted by considering the frequencies associated with the responses to a local pinch. However, since $\omega_*$ and $\omega_{\mbox{\tiny BP}}$ may differ~\cite{eric_boson_peak_emt}, it is not currently clear which of these frequencies is better represented by $\omega_g$.

The stiffening of QLMs by a factor of approximately $2$ accounts for an approximate $30$-fold variation of $A_g$, due to the $\omega_g^{-5}$ dependence in~\eqref{foo01}. The remaining variation is attributed to the number of QLMs, ${\cal N}\=A_g\,\omega_g^{5}$ (note that here we use an equality, as the $T_p$-independent prefactor is of no interest), plotted in Fig.~\ref{fig:Fig4}b. The result indicates that QLMs are depleted by slightly less than $2$ orders of magnitude in the simulated $T_p$ range. The strong depletion of QLMs upon deeper supercooling has dramatic consequences for the properties of the resulting glassy states. For example, brittle failure \cite{MW_yielding_2018_pre,Ozawa6656} and reduced fracture toughness \cite{Rycroft2012, Vasoya2016, Eran_mechanical_glass_transition} are claimed to be a consequence of this depletion. It is interesting to note that the range of variability observed in Fig.~\ref{fig:Fig4}b appears to be consistent with a very recent study~\cite{LB_two_level_systems_2019} of the depletion of tunneling two-level systems in stable computer glasses, possibly indicating that a subset of the QLMs is associated with tunneling two-level systems~\cite{soft_potential_model_1987,soft_potential_model_1991,SciPost2016,MW_theta_and_omega}.

The results presented in Fig.~\ref{fig:Fig4} demonstrate that pinching a glass may offer a procedure to separate the depletion and stiffening processes that take place with progressive supercooling. Next, we aim at exploring the physical implications of disentangling ${\cal N}$ and $\omega_g$.

\section{A thermodynamic signature of the QLMs}

QLMs correspond to compact zones (though they also have long-range elastic manifestations), which are embedded inside a glass, and characterized by particularly soft structures. It is tempting then to think of them as quasiparticles that feature well defined properties (e.g.~formation energy). If true, one may hypothesize that QLMs can be created and annihilated by thermodynamic fluctuations and follow an equilibrium distribution at the parent equilibrium temperatures $T_p$. Moreover, their equilibrium thermodynamic nature might be manifested in non-equilibrium glassy states as they become frozen in during the rapid quench upon glass formation.
\begin{figure}[ht]
\centering
\begin{tabular}{ccc}
\includegraphics[width=0.90\linewidth]{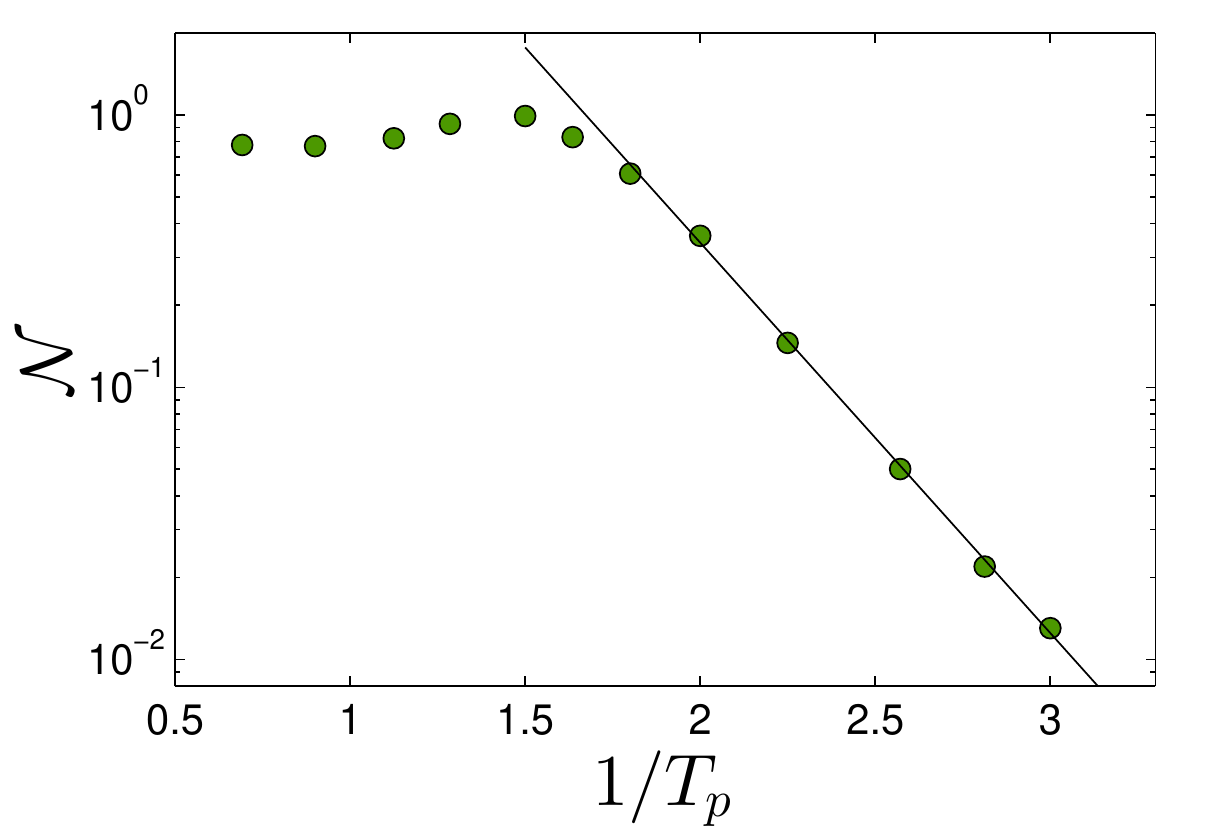}
\end{tabular}
\caption{The density of QLMs, plotted against $1/T_p$, revealing that it is controlled by a Boltzmann-like factor $e^{-E_{\mbox{\tiny QLM}}/{k_{\mbox{\tiny $B$}} T_p}}$, with the parent temperature playing the role of the equilibrium temperature. We find $E_{\mbox{\tiny QLM}}\!\approx\!3.3$, expressed in terms of $k_{\!_B} T_{\mbox{\tiny onset}}$.}
\label{fig:Fig5}
\end{figure}

As we have now at hand an estimate of the number ${\cal N}$ as a function of $T_p$ (cf.~Fig.~\ref{fig:Fig4}b), we can start testing these ideas. To this aim, we plot in Fig.~\ref{fig:Fig5} ${\cal N}$ vs.~$T_p^{-1}$ on a semilogarithmic scale; the outcome reveals a key result: the number of QLMs follows a Boltzmann-like law, with the \emph{parent temperature} $T_p$ playing the role of the equilibrium temperature, namely
\begin{equation}
{\cal N} \propto \exp\!\left(\!-\frac{E_{_{\mbox{\tiny QLM}}}}{k_{\!_B} T_p}\right)\,.
\label{eq:Boltzmann}
\end{equation}
A possibly related Boltzmann-like law, albeit for $A_g(T_p)$ itself, was observed in~\cite{mw_thermal_origin} for reheated stable glasses~\cite{fsp}. A corollary of~\eqref{eq:Boltzmann} is that QLMs seem to feature a well-defined formation energy, $E_{\mbox{\tiny QLM}}\!\approx\!3.3$ (in units of $k_BT_{\mbox{\tiny onset}}$). It is surprising that $E_{\mbox{\tiny QLM}}$ appears to be independent of $T_p$, while the characteristic energy scale associated with $\omega_g$ does appear to depend on it. Future research should shed additional light on this non-trivial observation.

The results in~\eqref{eq:Boltzmann} and Fig.~\ref{fig:Fig5} indicate that QLMs might indeed correspond to a subset of configurational degrees of freedom that equilibrate at the parent temperature $T_p$ and that carry memory of their equilibrium distribution when the glass goes out-of-equilibrium during a quench to lower temperatures. This physical picture strongly resembles the idea of a fictive/effective/configurational temperature, which was quite extensively used in models of the relaxation, flow and deformation of glasses~\cite{Tool1946, Narayanaswamy1971, Angell2000, Mauro2009fictive, Bouchbinder2009b, Bouchbinder2009c, Falk2011}. This connection is further strengthened in light of available evidence indicating that the cores of deformation-coupled QLMs are the loci of irreversible plastic events that occur once a glass is driven by external forces~\cite{lemaitre2004,micromechanics2016,zohar_prerc}.

The Boltzmann-like relation in~\eqref{eq:Boltzmann}, when interpreted in terms of STZs, is a cornerstone of the non-equilibrium thermodynamic STZ theory of the glassy deformation~\cite{Bouchbinder2009b, Bouchbinder2009c, Falk2011}, where $T_p$ is treated as a thermodynamic temperature that characterizes configurational degrees of freedom and that differs from the bath temperature. The strong depletion of STZs with decreasing $T_p$, as predicted by the Boltzmann-like relation, was shown to give rise to a ductile-to-brittle transition in the fracture toughness of glasses~\cite{Rycroft2012,Vasoya2016}. This prediction was recently supported by experiments on the toughness of Bulk Metallic Glasses (BMGs), where $T_p$ was carefully controlled and varied~\cite{Eran_mechanical_glass_transition}.

It is natural to define a length scale corresponding to the typical distance between QLMs as $\xi_s\!\sim\!{\cal N}^{-1/\dbar}$, once an estimate of their number ${\cal N}$ is at hand. Such a ``site length'' $\xi_s$ was introduced in~\cite{modes_prl}, where it was related to the sample-to-sample average minimal QLM frequency $\left<\omega_{\rm min}\right>$ according to $\left<\omega_{\rm min}\right>\!\sim\!\omega_g(L/\xi_s)^{-\dbar/5}$. The latter implies that the lowest QLM frequency is selected among $(L/\xi_s)^\dbar\!\propto\! {\cal N}\,N$ possible candidates, which is directly related to the extreme value statistics of $\omega_{\rm min}$~\cite{modes_prl}. The site length $\xi_s$ is expected to control finite-size effects in studies of athermal plasticity in stable glasses, as discussed in detail in \cite{Corrado_strain_intervals_pre_2018}. Similar definitions of a site length were proposed in \cite{karmakar_lengthscale,Corrado_strain_intervals_pre_2018}; an important message here is that the disentangling of the stiffening effect from the prefactor $A_g$ is imperative for the purpose of obtaining a consistent definition of a length scale in such a setting.

\section{A glassy length scale revealed by pinching a glass}

What additional physics can pinching a glass reveal? Up to now we explored the physics of the QLMs number ${\cal N}$; we now turn to the other contribution to $A_g$, i.e.~to the frequency scale $\omega_g$ that characterizes the typical stiffness of QLMs. $\omega_g$ was shown to undergo stiffening with decreasing $T_p$ (cf.~Fig.~\ref{fig:Fig4}a); is this stiffening related to other properties of QLMs that vary with $T_p$? An interesting possibility we explore here is whether it might be related to a glassy length scale that is associated with QLMs.

\begin{figure}[ht]
\centering
\begin{tabular}{ccc}
\includegraphics[width=0.875\linewidth]{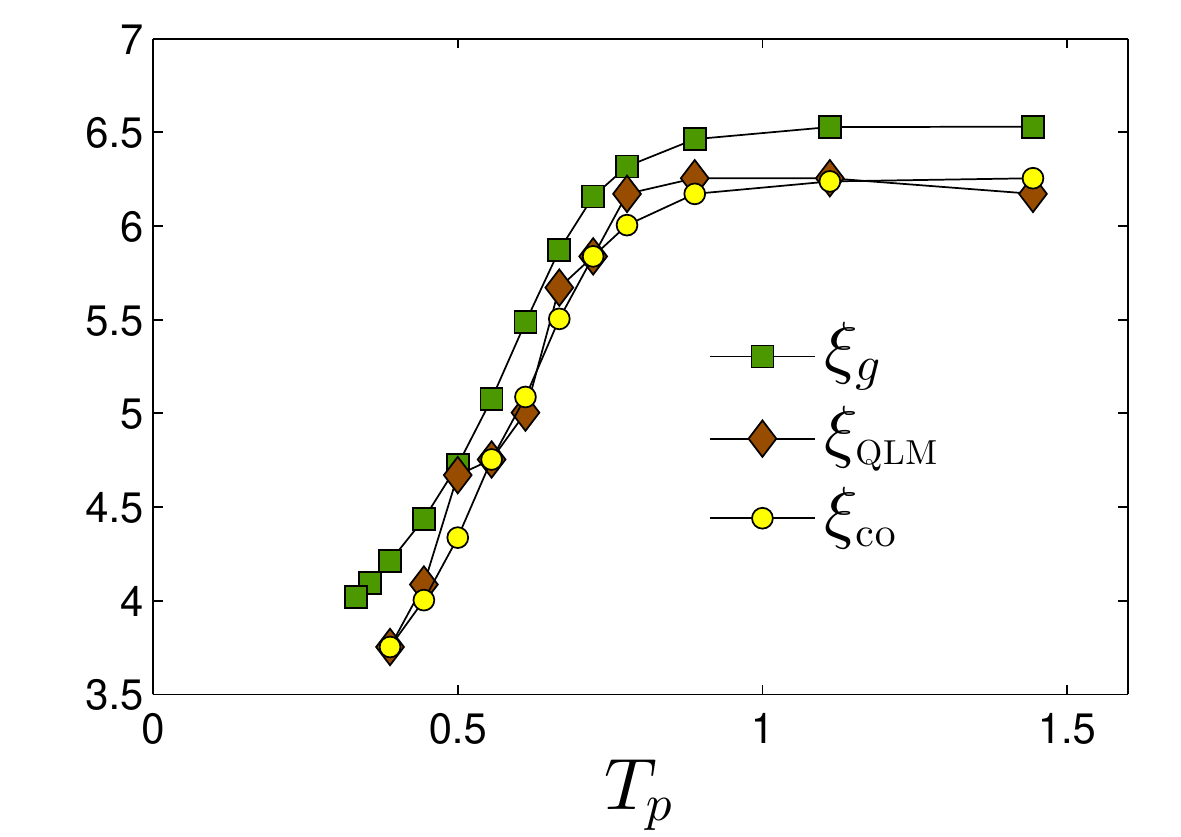}
\end{tabular}
\caption{(left) The glassy length $\xi_g$, the crossover length $\xi_{\mbox{\tiny co}}$, and the QLM core size $\xi_{\mbox{\tiny QLM}}$ (see SI for details), plotted against the parent temperature $T_p$. These lengths vary together with parent temperature $T_p$, supporting their equivalence.}
\label{fig:Fig6}
\end{figure}

To that aim, we construct a length scale $\xi_g$ as
\begin{equation}
\label{xi_g}
\xi_g \equiv 2\pi\,c_s/\omega_g\ ,
\end{equation}
which corresponds to the wavelength of transverse phonons propagating at the shear wave-speed $c_s$ with an angular frequency $\omega_g$. This length is similar in spirit to the ``boson peak'' length $\xi_{\mbox{\tiny BP}}\!\sim\!c_s/\omega_{\mbox{\tiny BP}}$~\cite{sokolov_boson_peak_scale}. The physical rationale behind our constructed length $\xi_g$ is that the emerging length scale is expected to mark a crossover in the elastic response of a glass to a local pinch, as discussed below. In Fig.~\ref{fig:Fig6} we plot $\xi_g$ vs.~the parent temperature $T_p$; we find that $\xi_g$ \emph{decreases} upon deeper annealing by approximately 40\%, a manifestation of the modest stiffening of the macroscopic shear modulus compared to that of QLMs (recall that $c_s$ is proportional to the square root of the shear modulus). This decreasing length is of unique character amongst the plethora of glassy lengthscales previously put forward in the context of the glass transition, most of which are increasing functions of decreasing temperature or parent temperature \cite{biroli_point_to_set_jcp_2004,Montanari2006,Biroli2008,glen_prl_2012_point_to_set,paddy_huge_review_2015,smarajit_review}.

In order to shed light on the physical meaning of $\xi_g$, we consider also $(i)$ the crossover length $\xi_{\mbox{\tiny co}}$ as observed in the displacement response to local pinches, between an atomistic-disorder-dominated response at distances $r\!\lesssim\!\xi_{\mbox{\tiny co}}$ from the perturbation, to the expected continuum behavior seen at $r\!>\!\xi_{\mbox{\tiny co}}$, and $(ii)$ the core size of QLMs, $\xi_{\mbox{\tiny QLM}}$, which is known to decrease upon annealing~\cite{protocol_prerc, inst_note, LB_modes_2019}, as is also illustrated graphically in Fig.~\ref{fig:Fig1}. In Fig.~\ref{fig:Fig6} we directly compare between $\xi_g$ and our measurements of  $\xi_{\mbox{\tiny co}}$ and $\xi_{\mbox{\tiny QLM}}$ (see SI for details). These three lengthscales feature very similar variations with $T_p$, strongly supporting their equivalence. Consequently, $\xi_g$ --- which was defined through the dipole response frequency $\omega_g$ (cf.~\eqref{xi_g}) --- seems to provide a measure of the core size of QLMs, and in light of the suggested relation between the latter and STZs, also of the size of STZs.

Additional insight may be gained by invoking the relation --- established in \cite{new_variational_argument_epl_2016} ---  between $\omega_g$ and the characteristic frequency $\omega_*$ that emerges near the unjamming transition \cite{ohern2003,liu_review,van_hecke_review}. Indeed, in the unjamming scenario the length $\xi_g$ (often denoted $\ell_c$) was shown to diverge upon approaching the unjamming point~\cite{breakdown}, and to mark the crossover between disorder-dominated responses near a local perturbation, and the continuum-like response observed in the far field, away from the perturbation. The same length was shown in~\cite{atsushi_core_size_pre} to characterize the core size of QLMs near the unjamming point of harmonic-sphere packings. In light of the results shown in Fig.~\ref{fig:Fig6}, we hypothesize that the fundamental crossover length --- below which responses to local perturbations are microstructural/disorder-dominated, and above which responses to local perturbations follow the expected continuum-like behavior --- is, in fact, $\xi_g$, which, in turn, we show to agree well with the size of QLMs.

\section{Summary and outlook}

In this work we have employed a computer glass model, which can be deeply annealed~\cite{LB_swap_prx}, to quantitatively study the variation of the properties of QLMs (soft spots) with the depth of annealing. Most notably, we calculated the variation of the number, characteristic frequency and core size of QLMs with the parent temperature from which the glass is formed. This has been achieved by assuming that the characteristic frequency scale of QLMs can be estimated through the bulk-average response of a glass to a local pinch. This frequency scale, in turn, allowed us to disentangle the apparently inseparable effects of the depletion and stiffening of QLMs, which are both encoded in the prefactor of the universal $\omega^4$ vibrational density of states of QLMs.

We found that the number of QLMs follows a Boltzmann-like factor, with the parent temperature --- from which equilibrium configurations were vitrified --- playing the role of the equilibrium temperature. Consequently, the parent temperature may be regarded as a non-equilibrium temperature that characterizes QLMs deep inside the glassy state. Furthermore, our analysis reveals that both the core size of QLMS, and the mesoscopic length scale that marks the crossover between atomistic-disorder-dominated responses near local perturbations, and continuum like responses far away from local perturbations, can be estimated using the characteristic frequency of QLMs --- obtained by pinching the glass ---, and the speed of shear waves.

Our results may have important implications for various basic problems in glass physics. We mention a few of them here; first, the Boltzmann-like law of the number of QLMs may play a major role in theories of the relaxation, flow and deformation of glasses, and may support some existing approaches. Second, together with other available observations~\cite{new_variational_argument_epl_2016,LB_modes_2019,atsushi_core_size_pre}, our results may suggest that the boson peak frequency could be robustly probed by pinching glassy samples, instead of the more involved analysis required otherwise~\cite{Marruzzo2013,LB_modes_2019}. Finally, the variation of the \emph{energy scale} proportional to $\omega_g^2$ with annealing temperature appears to match very well the variation of activation barriers required to rationalize fragility measurements in laboratory glasses (compare Fig.~\ref{fig:Fig4}a with Fig.~8 of~\cite{tarjus_2004}). If valid, our results appear to support elasticity-based theories of the glass transition~\cite{elastic_model2006,swap_prx_MW,MW_cates_length_discussion_prl_2017}, and indicate that QLMs play important roles in relaxation processes in deeply supercooled liquids~\cite{widmer2008irreversible}. We hope that these interesting investigation directions will be pursued in the near future.

\appendix

\section{Models and methods}
We employed a computer glass forming model in three dimensions, simulated using the swap Monte-Carlo method, explained e.g.~in \cite{LB_swap_prx}. The model consists of soft repulsive spheres interacting via a $\propto r^{-10}$ pairwise potential (with $r$ denoting the distance between the centers of a pair of particles), enclosed in a fixed-volume box with periodic boundary conditions. The particles' sizes are drawn from a distribution designed such that crystallization is avoided \cite{LB_swap_prx}. A comprehensive description of the model, and of all parameter choices, can be found in \cite{boring_paper}, including an important discussion about how we handled large sample-to-sample realization fluctuations of particle sizes that can arise in small system sizes due to the breadth of the employed particle size distribution. Ensembles of 10000, 1000, and 2000 glassy samples were made for systems of $N\!=\!2000,8000,$ and 16000 particles, respectively, by instantaneously-quenching (to zero temperature) independent configurations equilibrated at various parent temperatures $T_p$. All data except for those shown in Fig.~\ref{fig:Fig6} were calculated using the smaller glasses. Lengths are expressed in terms of $a_0\!\equiv\! (V/N)^{1/\dbar}$ where $V$ is the system's volume. All particles share the same mass $m$, which we set as our microscopic unit of mass. Frequencies are expressed in terms of $c_\infty/a_0$, where $c_\infty\!\equiv\!\sqrt{G_\infty/\rho}$ is the high-$T_p$ shear wave-speed, with $G_\infty$ denoting the high-$T_p$ sample-to-sample mean athermal shear modulus, and $\rho\!\equiv\! mN/V$ denotes the mass density. $T_p$ is expressed in terms of the crossover temperature $T_{\mbox{\tiny onset}}$, above which the sample-to-sample mean athermal shear modulus saturates to a high-temperature plateau, as shown in the inset of Fig.~\ref{fig:Fig4}a and in \cite{boring_paper}. In our model we find $G_\infty a_0^3/k_BT_{\mbox{\tiny onset}}\!\approx\! 17$. Data will be made available upon request from the corresponding author.

\vspace{0.3cm}

\acknowledgements

We thank David Richard for his help with our graphics. Fruitful discussions with David Richard and Geert Kapteijns are warmly acknowledged. E.~B.~acknowledges support from the Minerva Foundation with funding from the Federal German Ministry for Education and Research, the Ben May Center for Chemical Theory and Computation, and the Harold Perlman Family. E.~L.~acknowledges support from the Netherlands Organisation for Scientific Research (NWO) (Vidi grant no.~680-47-554/3259).


%


\newpage

\onecolumngrid
\begin{center}
	\textbf{\large Supporting Information for: ``What can be learnt from pinching a glass?''}
\end{center}

\setcounter{equation}{0}
\setcounter{figure}{0}
\setcounter{section}{0}
\setcounter{table}{0}
\setcounter{page}{1}
\makeatletter
\renewcommand{\theequation}{S\arabic{equation}}
\renewcommand{\thefigure}{S\arabic{figure}}
\renewcommand{\thesubsection}{S-\arabic{subsection}}
\renewcommand{\thesection}{S-\arabic{section}}
\renewcommand*{\thepage}{S\arabic{page}}
\renewcommand{\bibnumfmt}[1]{[S#1]}
\renewcommand{\citenumfont}[1]{S#1}
\twocolumngrid

In this Supporting Information we provide information about ($i$) how the bulk-average frequency of the response to a local pinch of the glass, denoted $\omega_g$ in the manuscript, was calculated, ($ii$) how the modes in Fig.~1 of the main text were calculated, and their size estimated, and ($iii$) how we estimated the crossover length $\xi_{\mbox{\tiny co}}$ and the QLMs core size $\xi_{\mbox{\tiny QLM}}$, both appearing in Fig.~6 of the main text.

We recall that lengths are expressed in terms of $a_0\!\equiv\! V/N$ where $V$ is the system's volume, and $N$ denotes the number of particles. All particles share the same mass $m$, which we set as our microscopic unit of mass. Frequencies are expressed in terms of $c_\infty/a_0$, where $c_\infty\!\equiv\!\sqrt{G_\infty/\rho}$ is the high-$T_p$ shear wave-speed, with $G_\infty$ denoting the high-$T_p$ plateau of sample-to-sample mean athermal shear modulus of inherent states (see inset of Fig.~3a of main text), and $\rho\!\equiv\! mN/V$ denotes the mass density. Temperatures are expressed in terms of the crossover temperature $T_{\mbox{\tiny onset}}$, above which the sample-to-sample mean athermal shear modulus saturates to a high-temperature plateau, as seen in Fig.~4 of the main text, and in \cite{boring_paper}. In our system we find $G_\infty a_0^3/k_BT_{\mbox{\tiny onset}}\!\approx\! 17$, with $k_B$ denoting the Boltzmann constant.

\subsection{The calculation of $\omega_g(T_p)$}

Following \cite{cge_paper_}, we define a local dipolar force as
\begin{equation}
\dv^{(ij)}_k \equiv \frac{\partial\varphi_{ij}}{\partial\xv_k}\,,
\end{equation}
where Roman indices denote particle indices, $\varphi_{ij}$ is the radially-symmetric pairwise potential between the $i^{\mbox{\tiny th}}$ and $j^{\mbox{\tiny th}}$ particles, and $\xv_k$ denotes the $\dbar$-dimensional coordinate vector of the $k^{\mbox{\tiny th}}$ particle. The linear displacement response to this force dipole reads
\begin{equation}\label{moo01}
\uv_k^{(ij)} = {\calBold H}^{-1}_{k\ell}\cdot\dv_\ell^{(ij)}\,,
\end{equation}
where repeated indices are understood to be summed over, and
\begin{equation}
{\calBold H}_{k\ell}\equiv\frac{\partial^2U}{\partial\xv_k\partial\xv_\ell}
\end{equation}
is the Hessian of the potential energy $U\!\equiv\!\sum_{i<j}\varphi_{ij}$. The frequency associated with the response $\uv$ is given by
\begin{equation}
\omega_g^{(ij)}\equiv \frac{1}{\sqrt{m}}\left( \frac{\uv^{(ij)}_k\cdot{\calBold H}_{k\ell}\cdot\uv^{(ij)}_\ell}{\uv^{(ij)}_k\cdot\uv^{(ij)}_k}\right)^{1/2}\,,
\end{equation}
where $m$ denotes the microscopic units of mass.

We next define the conditional average
\begin{equation}
\omega_g \equiv \langle \omega_g^{(ij)} \rangle_{f_{ij}/(pa_0^2)<10^{^{-2}}}\,,
\end{equation}
where $p$ is the glass pressure (recall that in our computer glass particles interact via a purely repulsive pairwise interaction), and the triangular brackets denote an average taken over all interacting pairs $i,j$ for which the dimensionless pairwise force $f_{ij}/(pa_0^2)\!<\!10^{-2}$.

\begin{figure}[!h]
\centering
\includegraphics[width = 0.49\textwidth]{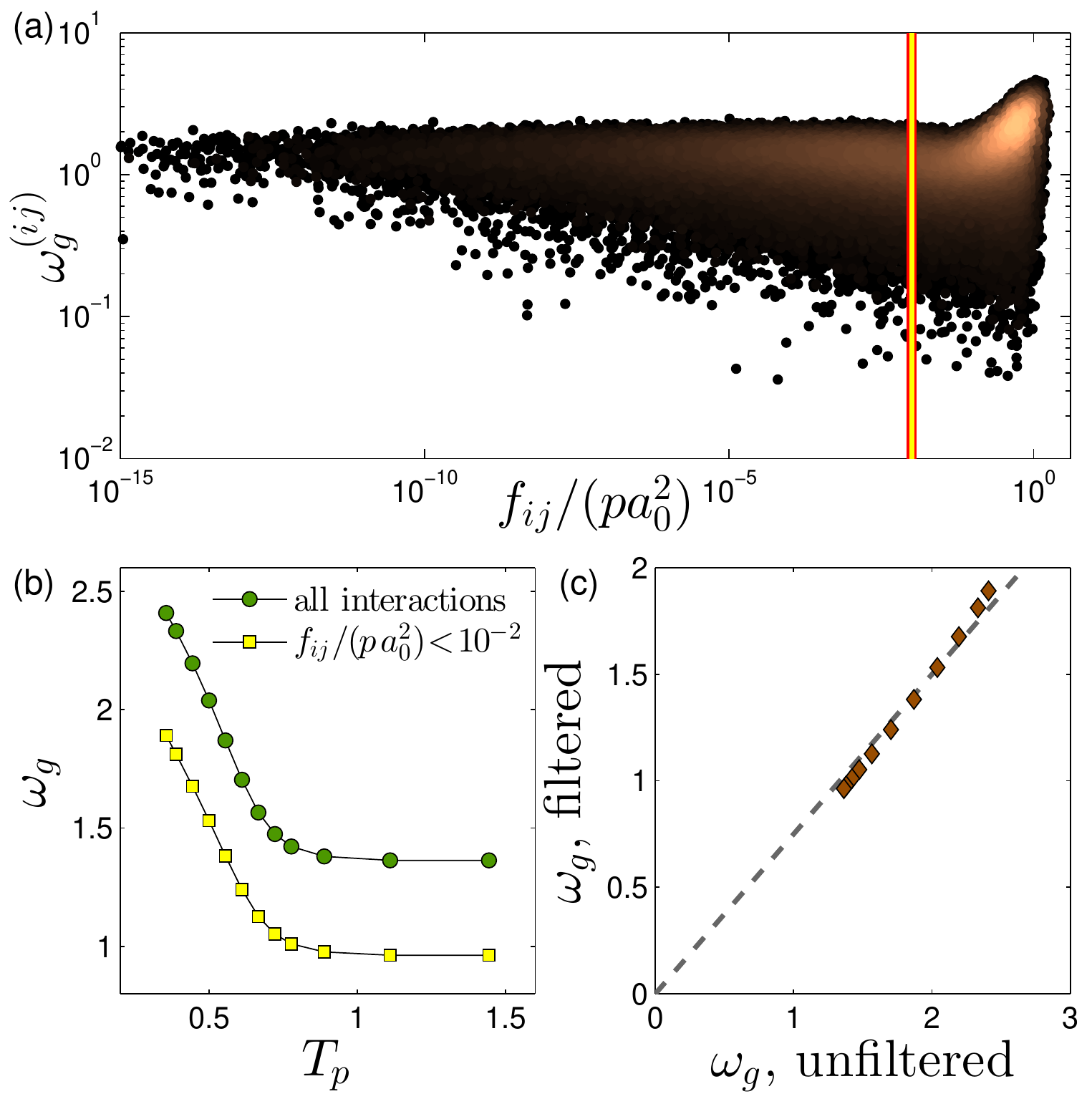}
\caption{\footnotesize (a) Scatter plot of $\omega_g^{(ij)}$ vs.~the dimensionless pairwise force $f_{ij}/(pa_0^2)$, calculated in an ensemble of glassy solids quenched from $T_p\!=\!5/9$. Results for other parent temperatures have similar forms. At strong forces the stiffnesses associated with responses to local pinches are substantially higher compared to those associated with weak forces. We find a saturation of the statistics of $\omega_g^{(ij)}$ at $f_{ij}/(pa_0^2)\!\lesssim\!10^{-2}$, marked by the vertical line. (b) \& (c) Comparison between the mean frequencies $\omega_g$ calculated with and without filtering by the pairwise forces; the two means differ by $\approx\!40\%$ consistently throughout the sampled temperature range. }
\label{fig:scatter_plot}
\end{figure}

The reason we chose to only consider weak forces in the estimation of $\omega_g$ can be understood by scatter-plotting $\omega_g^{(ij)}$ vs.~$f_{ij}/(pa_0^2)$, as seen in Fig.~\ref{fig:scatter_plot}a. We can clearly see that two families of frequencies are generated by pinching pairs between which strong or weak forces are found. In particular, strongly-interacting pairs tend to generate much stiffer responses (note the logarithmic $y$-axis). Since these responses are supposed to represent soft, quasilocalized modes, we opt for filtering the responses according to the dimensionless forces $f_{ij}/(pa_0^2)$. Below the chosen threshold $f_{ij}/(pa_0^2)<10^{-2}$, that can be clearly read off the scatter plot Fig.~\ref{fig:scatter_plot} (vertical yellow line), the statistics of $\omega_g^{(ij)}$ appears to saturate.

In Fig.~\ref{fig:scatter_plot}b,c we examine the effect of filtering interactions by their force on the $T_p$ dependence of $\omega_g$. We see that the relative variation of the two mean frequencies is very similar throughout the sampled parent-temperature range.

\subsection{Calculation of soft modes in 2D}
\label{soft_modes_2D}
In this Section we describe how the modes shown in Fig.~1 of the main text were calculated. A detailed description of this calculation will be presented elsewhere~\cite{in_prep_}, and see also \cite{kapteijns2019nonlinear_}; here the main points are summarized.

We employed the two-dimensional version of the same computer glass model used for our study; details about the model can be found in \cite{scattering_jcp}. Ensembles of glassy samples were quenched from equilibrium parent temperatures of $T_p\!=\!7/9$ (expressed in terms of the onset temperature $T_{\mbox{\tiny onset}}$ as described above) and $T_p\!=17/90$. We followed the framework put forward in \cite{Scipost2016_,kapteijns2019nonlinear_}, and calculated solutions $\calBold{\pi}$ to the equation
\begin{equation}\label{moo00}
{\calBold H}\cdot\calBold{\pi} = \frac{\calBold{\pi}\cdot{\calBold H}\cdot\calBold{\pi}}{\frac{\partial^4U}{\partial\xv\partial\xv\partial\xv\partial\xv}::\calBold{\pi}\calBold{\pi}\calBold{\pi}\calBold{\pi}}\frac{\partial^4U}{\partial\xv\partial\xv\partial\xv\partial\xv}:\!\!\cdot\,\calBold{\pi}\calBold{\pi}\calBold{\pi}\,,
\end{equation}
where triple and quadruple contractions are denoted as $:\!\!\cdot$ and $::$, respectively, and particle indices were suppressed for simplicity. Solutions $\calBold{\pi}$ to Eq.~(\ref{moo00}) were coined `quartic modes' \cite{cge_paper_,kapteijns2019nonlinear_}; they represent soft quasilocalized excitations that resemble low-frequency quasilocalized vibrational modes seen below or in between phonon bands \cite{Scipost2016_}, i.e.~in the absence of hybridizations with phonons. Solutions to Eq.~(\ref{moo00}) were calculated by employing a standard nonlinear conjugate gradient minimization algorithm to find local minima of the cost function \cite{Scipost2016_}
\begin{equation}\label{moo02}
{\cal G}(\zv) \equiv \frac{(\zv\cdot{\calBold H}\cdot\zv)^2}{\frac{\partial^4U}{\partial\xv\partial\xv\partial\xv\partial\xv}::\zv\zv\zv\zv}\,,
\end{equation}
where $\zv$ represents a displacement field in the $N\!\times\!\dbar$ dimensional configuration space of the glass. It is straightforward to show (see further details in \cite{Scipost2016_}) that minima of ${\cal G}$ correspond to solutions of Eq.~(\ref{moo00}). Initial conditions for the minimization of ${\cal G}$ were obtained by calculating the linear displacement response to a dipolar force, as given by Eq.~(\ref{moo01}), for every pair of interacting particles in the glass. In Fig.~1 of the main text, we only show modes $\calBold{\pi}$ for which $\omega_{\mbox{\scriptsize$\calBold{\pi}$}} \!\equiv\! \sqrt{\calBold{\pi}\!\cdot\! \calBold{H}\!\cdot\!\calBold{\pi}}/\sqrt{m}\!<\!\omega_g/3$, where $\omega_g$ was calculated as described in the previous Section. The area of the disordered core of each of the calculated modes was estimated as $Ne$, with $e$ denoting the participation ratio of the modes, defined for a normalized mode $\hat{\zv}$ as $e\!=\!\big(N\sum_i (\hat{\zv}_i\!\cdot\!\hat{\zv}_i)^2\big)^{-1}$. The participation ratio is a proxy for the degree of localization of a mode; in particular, for a localized mode one expects $e\!\sim\!{\cal O}(1/N)$, whereas a spatially-extended mode would give $e\!\sim\!{\cal O}(1)$. In Fig.~1 of the main text, the area of each blob that represents a soft mode is proportional to $Ne$, and its color represents its frequency $\omega_{\mbox{\scriptsize$\calBold{\pi}$}}$, with dark (bright) colors representing softer (stiffer) modes.

\subsection{Estimation of the crossover length}
In this Section we describe how we measured the crossover length $\xi_{\mbox{\tiny co}}$ between disorder-dominated responses near a local perturbation, to continuum, Eshelby-like algebraic decays away from a local perturbation. The crossover lengths $\xi_{\mbox{\tiny co}}$ extracted from the following analysis are shown in Fig.~6 of the main text.

\begin{figure}[!h]
\centering
\includegraphics[width = 0.5\textwidth]{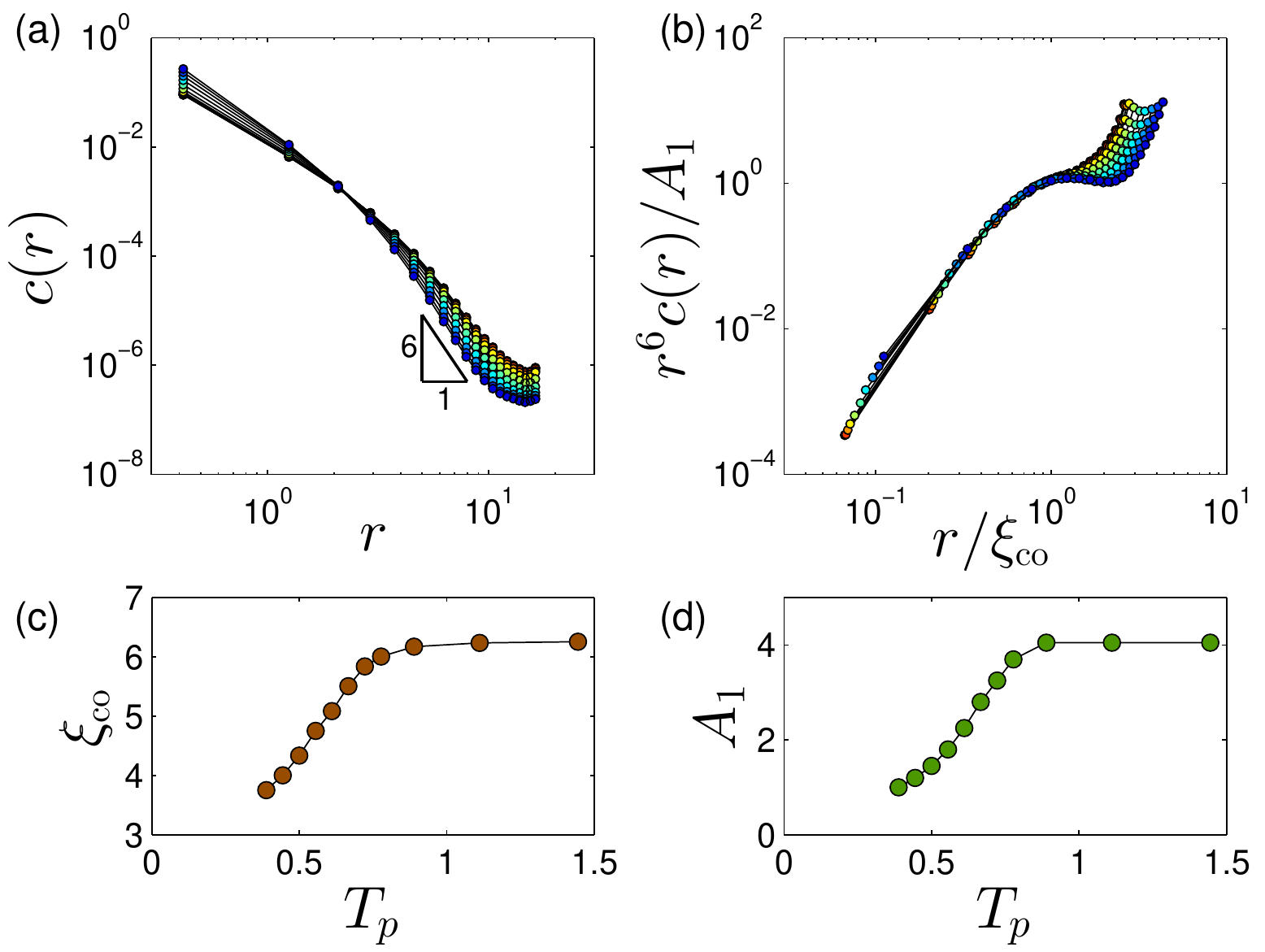}
\caption{\footnotesize (a) Decay functions $c(r)$ of the response to local pinches, see text for precise definition. The different curves correspond to measurements performed on glassy samples quenched from $T_p\!=\!13/9,10/9,8/9,7/9,13/18,2/3,11/18,5/9, 1/2, 4/9, 7/18$, from warm to cold colors. (b) Rescaling $c(r)$ by $r^{-6}$ allows to robustly identify the crossover length $\xi_{\mbox{\tiny co}}$ between disorder-dominated to continuum-like scaling. (c) \& (d) The crossover lengths $\xi_{\mbox{\tiny co}}$ and the factors $A_1$ used to collapse the curves in panel (b), plotted against the parent temperature $T_p$.}
\label{fig:dipole_response_decay}
\end{figure}

In order to estimate the crossover length, we follow the measurement scheme of \cite{breakdown_}; this amounts to calculating the response to local dipoles via Eq.~(\ref{moo01}), still following the dimensionless-force filtering scheme discussed above. The fields are then normalized, namely for every pair $ij$ considered, we calculate $\hat{\uv}^{(ij)}\!\equiv\!\uv^{(ij)}/|\uv^{(ij)}|$. Then, for \emph{each} interaction $k\ell\!\ne\! ij$, we compute the square of the projection of the normalized response $\hat{\uv}^{(ij)}$ onto the normalized dipole vector $\hat{\dv}^{(k\ell)}\!\equiv\!\dv^{(k\ell)}/|\dv^{(k\ell)}|$, i.e.~we calculate
\begin{equation}
C_{ij,k\ell} \equiv \big( \hat{\uv}^{(ij)}\cdot\hat{\dv}^{(k\ell)} \big)^2\,.
\end{equation}
$C_{ij,k\ell}$ generally depends on the distance $r_{ij,k\ell}$ between the interactions $ij$ and $k\ell$, and on their relative orientation.

For each normalized response field $\hat{\uv}^{(ij)}$, we bin $C_{ij,k\ell}$ --- calculated for all pairs $k\ell\!\ne\! ij$ --- over the distances $r_{ij,k\ell}$, and calculate the \emph{median} of $C_{ij,k\ell}$ over all pairs $k\ell$ located at similar distances $r$ away from the excited dipole $\dv^{ij}$; the average over the excited dipole $ij$, and over glassy samples, denoted below by $\langle\bullet\rangle_{ij}$, defines the decay function $c(r)$, namely
\begin{equation}
c(r) \equiv \big< \underset{r_{ij,k\ell}\approx r}{\mbox{median}_{k\ell}}\big( C_{ij,k\ell} \big) \big>_{ij}\,.
\end{equation}
The decay functions $c(r)$ are plotted in Fig.~\ref{fig:dipole_response_decay}a. Continuum elasticity would predict that $c(r)\!\sim\! r^{-2\dbar}$ \cite{breakdown_}. We therefore plot in Fig.~\ref{fig:dipole_response_decay}b the rescaled decay functions $r^6c(r)/A_1$ against the rescaled distance $r/\xi_{\mbox{\tiny co}}$ with $\xi_{\mbox{\tiny co}}(T_p)$ denoting the parent-temperature dependent crossover lengths, chosen to collapse the data, as are the constants $A_1(T_p)$ reported in Fig.~\ref{fig:dipole_response_decay}d. The crossover lengths $\xi_{\mbox{\tiny co}}(T_p)$ are plotted against the parent temperature $T_p$ in Fig.~\ref{fig:dipole_response_decay}c.


\subsection{Estimation of QLMs core length}
\begin{figure}[!h]
\centering
\includegraphics[width = 0.5\textwidth]{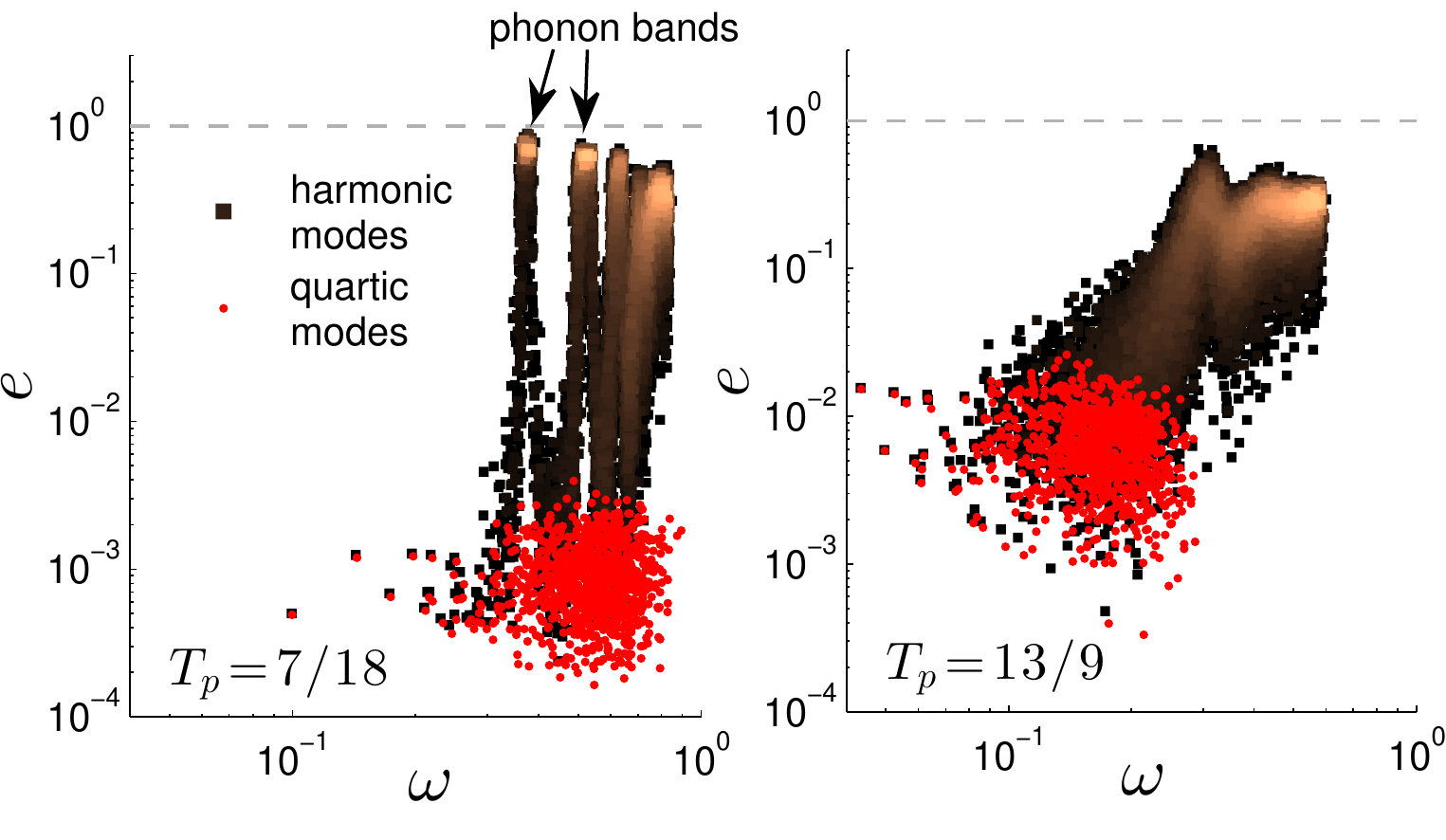}
\caption{\footnotesize Scatter plot of the participation ratio $e$ --- that quantifies the degree of localization of a mode --- vs.~frequency, calculated for harmonic (black symbols) and quartic (red symbols) modes in glassy samples of $N\!=\!8000$ particles quenched from $T_p\!=\!7/18$ (left) and $T_p\!=\!13/9$ (right). In the analyzed stable glassy samples ($T_p\!=\!7/18$, left panel) phonons and QLMs dwell at similar frequencies, leading to their hybridizations. Quartic modes appear to be entirely indifferent to the presence of these phonons (see left panel).}
\label{fig:scatter_participation_vs_freq}
\end{figure}

In stable glasses, it becomes difficult to sample many QLMs using a harmonic analysis due to their stiffening and depletion, discussed in length the main text. As a result of these processes, characteristic frequencies of the softest QLMs tend to overlap with the lowest phonon frequencies, leading to hybridizations of phonons and QLMs, and obscuring a clear picture of QLMs properties and statistics, as demonstrated in Fig.~\ref{fig:scatter_participation_vs_freq}.

In order to reveal the properties of QLMs for glasses quenched from all parent temperatures, including in stable glasses, we opt for calculating `quartic modes' as representatives of QLMs, since the former are known to be indifferent to the presence of phonons with comparable frequencies (they show no hybridizations with phonons, as shown in \cite{Scipost2016_} and in the left panel of Fig.~\ref{fig:scatter_participation_vs_freq}). At the same time, quartic modes feature frequencies that are in excellent agreement with QLMs' frequencies in the absence of hybridizations \cite{Scipost2016_}, as can also be seen in Fig.~\ref{fig:scatter_participation_vs_freq}.

We first generated, for each of our glassy samples of $N\!=\!8000$ particles, a quartic mode as discussed in length in Sect.~\ref{soft_modes_2D}. In this case, however, the initial conditions for finding quartic modes were chosen to be the linear displacement responses to the forces that arise due to imposing simple and pure shear \cite{lemaitre2006_avalanches_} in all possible Cartesian planes (i.e.~$\hat{x}\!-\!\hat{y},\hat{x}\!-\!\hat{z},\ldots$). Each such linear response is then used as the initial condition for the minimization of the cost function ${\cal G}$, c.f.~Eq.~(\ref{moo02}). An ensemble of QLMs is constructed by only keeping and considering the QLM $\mathBold{\pi}$ with the smallest frequency $\omega_{\mbox{\scriptsize$\calBold{\pi}$}} \!\equiv\! \sqrt{\calBold{\pi}\!\cdot\!\calBold{H}\!\cdot\!\calBold{\pi}}/\sqrt{m}$ amongst all those calculated \emph{for each individual sample}, leaving us with 1000 soft QLMs per parent temperature~$T_p$.

In order to demonstrate the utility of quartic modes for the assessment of the core size of QLMs, we scatter-plot in Fig.~\ref{fig:scatter_participation_vs_freq} the participation ratio of both harmonic modes (obtained by a partial diagonalization of the Hessian of the potential energy), and quartic modes (obtained as described above). We show that, at the very lowest frequencies, each harmonic mode overlaps with a quartic mode that our calculation produces, demonstrating that our calculation captures well the QLM away from regimes of strong hybridizations with phonons. These data show that harmonic and quartic mode share very similar localization properties and frequencies, as also discussed in length in \cite{Scipost2016_}, which motivates employing quartic modes as faithful representitives of QLM.

\begin{figure}[!h]
\centering
\includegraphics[width = 0.5\textwidth]{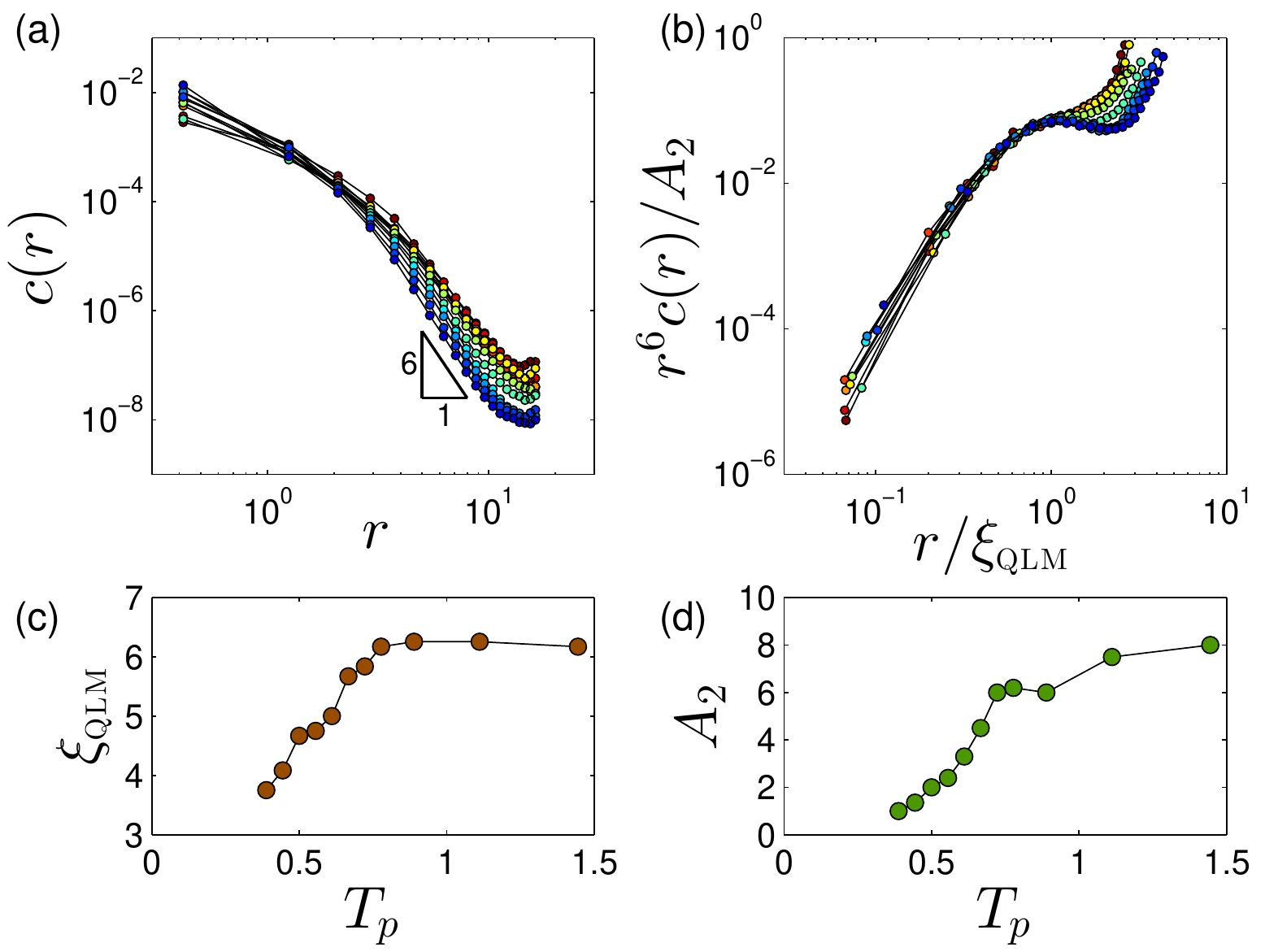}
\caption{\footnotesize (a) Decay functions $c(r)$ of QLMs calculated as explained in this SI. The different curves correspond to measurements performed on glassy samples quenched from the same parent temperatures $T_p$ as spelled out in the caption of Fig.~\ref{fig:dipole_response_decay}. (b) Rescaling $c(r)$ by $r^{-6}$ allows to robustly identify the QLMs linear core size $\xi_{\mbox{\tiny QLM}}$. (c) \& (d) The QLM core length $\xi_{\mbox{\tiny QLM}}$ and the factors $A_2$ used to collapse the curves in panel (b), plotted against the parent temperature $T_p$.}
\label{fig:qlm_size_fig_si}
\end{figure}

In order to estimate the linear size of the cores of QLM, each calculated QLM $\mathBold{\pi}$ as described above was normalized $\hat{\mathBold{\pi}}\!\equiv\!\mathBold{\pi}/|\mathBold{\pi}|$; we then identified the pair $ij$ of interacting particles that maximizes the difference squared $|\hat{\mathBold{\pi}}_i\!-\hat{\mathBold{\pi}}_j|^2$, and consider this pair as the center of the QLM's core. We calculated the spatial decay $c(r)$ of QLMs similarly to the procedure explained in the previous Section for analyzing the spatial decay of the response to a local pinch, with the only differences being that here that $r$ represents the distance from the aformentioned pair $ij$,
\begin{equation}
C_{ij,k\ell} \equiv \big( \hat{\mathBold{\pi}}\cdot\hat{\dv}^{(k\ell)} \big)^2\,,
\end{equation}
and
\begin{equation}
c(r) \equiv \big< \underset{r_{ij,k\ell}\approx r}{\mbox{median}_{k\ell}}\big( C_{ij,k\ell} \big) \big>_{\mbox{\tiny QLMs}}\,,
\end{equation}
where the average is taken over all calculated QLMs.

The results of this calculation are shown in Fig.~\ref{fig:qlm_size_fig_si}, see figure caption for further details. The lengths $\xi_{\mbox{\tiny QLM}}$ extracted from our analysis are shown in Fig.~\ref{fig:qlm_size_fig_si}c, and used in Fig.~6 of the main text.

\end{document}